\documentclass[12pt]{article}
\usepackage{graphicx}
\usepackage{lscape}
\bibstyle{plain}

\newcommand{\be}{\begin{equation}}
\newcommand{\en}{\end{equation}}

\newcommand{\PP}{{\mathord{I\kern -.33em P}}}
\newcommand{\EE}{{\mathord{I\kern -.33em E}}}
\newcommand{\RR}{{\mathord{I\kern -.33em R}}}

\newcommand{\ea}{\end{eqnarray}}
\newcommand{\ba}{\begin{eqneduardarray}}
\newcommand{\ean}{\end{eqnarray*}}
\newcommand{\ban}{\begin{eqnarray*}}

\begin{document}

\title{The role of the Model Validation function to manage and mitigate model risk}

\author{Alberto Elices\thanks{Head of Equity Model Validation, Risk Methodology, Divisi\'{o}n General de Riesgos,  Santander, Ciudad Financiera Santander, Avda. Cantabria s/n, 28660 Boadilla del Monte, Spain, {\em aelices@gruposantander.com}.}}
\date{\today}
\maketitle

\bigskip

\begin{abstract}
This paper describes the current taxonomy of model risk, ways for its mitigation and management and the importance of the model validation function in collaboration with other departments to design and implement them.
\end{abstract}

\section{Introduction}
\label{sec:Introduction}

After the start of the crisis in 2008, a big concern about pricing models has been raised. Risk management and model validation have drawn considerably more attention. From the model validation perspective, implementation testing is no longer enough and risk department is more concerned about model limitations and involved to control model risk exposure.

Pricing models are evolving toward simpler products with more complex model assumptions. Therefore, the current trend is more focused on improving existing models rather than developing new ones. Modelling improvement is not only about developing just a ``good'' model but also about reviewing basic model assumptions and about how the model integrates in the whole pricing system by improving the inputs (quality and consistency of market data) and getting appropriate and stable outputs for hedging and risk management purposes.

The validation function comes into play to foster and guarantee this constant improvement of pricing models by defining appropriate policies for mitigation and management of model risk through valuation control and fair value adjustment. This task should be done in collaboration and in good working relationship with other departments such as Market Risk or Front Office. This will ensure knowing the actual impact of these policies in the P\&L and make senior management and model users aware of it so that they may express how much appetite they are willing to stand.

\section{Taxonomy of model sources of uncertainty}
\label{sec:Taxonomy}

Model uncertain or unexpected behaviour may arise from many different sources: bad implementation, wrong use of model, uncertain model parameters, difficulty to obtain consistent market data, evolution of market consensus or missing key sources of risk like considering a stochastic factor to be deterministic. Some of these sources of uncertainty can be filtered out in the validation process such as bad implementation, wrong use of model, identification of missing sources of risk or identification of models which do not follow market consensus.

However, there are other sources of uncertainty which cannot be avoided after validation, even for a model which is not possible to improve in practice. For instance, uncertain model parameters, illiquid low quality market data or simplified modelling of certain risk factors. In some situations, there is no other solution to do business that the use of limited models with controllable risk. These sources of model uncertainty which cannot be avoided through the validation process and are not related with uncertainty of market data is what is commonly called ``model risk''.

For fixed income markets, basic model assumptions such as lognormality of interest rates or multi-curve modelling are being reviewed. Other examples of sources of model risk may be the auto-correlation amongst fixed income indexes (e.g. modelling the distribution of forward libor, constant maturity and regular swap indexes or deferred Libor payments), lack of modelling of stochastic basis, calibration risk or volatility modelling (construction, interpolation and extrapolation of implied volatility surface). Other sources of risk come from estimation of market data (correlations between interest rates and credit or correlations for quanto adjustments) or cross-sensitivity risk like for instance vanna found in notional increasing or accreting bermudan swaptions.

In equity markets, the current examples of model risk mainly arise from forward skew modelling (using stochastic local volatility models for cliquet, barrier and autocallable products), impact of stochastic interest rates for barrier and autocallable products, multi-curve modelling, marking correlation level and considering correlation skew. Other sources of risk come out of the hedging management of digital, callable and barrier payments mitigated with maximum delta and gamma softening\footnote{By maximum delta or gamma softening, it is meant that the payoff function is changed so that the maximum slope for delta and the maximum change of slope for gamma is limited to specified levels.}. Hedging risks also arise with cliquet options with local cap and floor (mitigated by maximum gamma softening), high cross-gamma baskets with deltas changing too much with component movements or very skew-dependent products whose hedging is considerably improved using vega maps instead of vega term structure. In addition, sources of risk coming from estimation of market data such as dividends or correlation for quanto and composite options may also be considered.

Inflation markets show model risk for instance from the lack of considering volatility smile, uncertainty around modelling the correlation structure among CPI (Consumer Price Index) rates and the correlation between inflation and interest rates. 

Foreign exchange markets show model risk from forward skew modelling for barrier options coming from the estimation of the stochastic parameters of stochastic local volatility models. Also the impact of stochastic interest rates and correlation modelling for multi-asset products, where there is uncertainty about correlations not derived from at-the-money volatilities as they depend on the not-well-solved problem of volatility surface construction for illiquid pairs.

\section{Managing and mitigating model risk}
\label{sec:ModelRisk}

Management of model risk is a combination of qualitative and quantitative assessments. Ideally, quantitative metrics are always preferred. However, when they are not available, at least a qualitative classification is necessary to make model users and senior management aware about model risk.

On the qualitative measures, management of model risk should start by periodically reviewing pricing models. This involves a complete inventory and classification of products with the models and engines which should be used to price them and decommission old ones. On a second stage, a control policy should be designed on the use of market data for calibration and the non-calibrated valuation parameters. This valuation control policy should also monitor difference between production models and market, either comparing with benchmark prices or trying to study cases of collateral dispute. On a third stage, policies to mitigate sources of uncertainty and to calculate fair value adjustments (FVA) should be designed and implemented. Mitigation policies can be for instance setting conservative values for internal parameters such as correlations, monitoring and setting limits to certain risks such as sensitivity to correlation, cross gamma or vanna, reduce hedging risk by maximum delta or gamma softening or limit product features like forward start terms or deal maturities.

On the quantitative side, fair value adjustment allows quantifying model risk. It can be defined as a metric to measure model uncertainty (see \cite{Cont2006} or \cite{Morini2011}). It should cover the expected hedging loss of a given portfolio plus some of the uncertainty of that loss (see \cite{Elices2011}). The expected benefit of an operation (client price minus market price) which can be realized on a given date is equal to the total benefit minus the FVA (taken out to provide some cushion in case things go wrong). Good qualities of a good FVA policy are: it should be dynamic, stable, transparent, easy to compute and decrease as uncertainty gets reduced (usually as expiration approaches). It should also foster the improvement of models by reducing the FVA when the model progressively gets improved.

Fair value adjustments can be calculated in many ways. For instance, the uncertainty of a given non-calibrated parameter (correlation, dividends, mean reversion) can be captured by looking at the portfolio price variation when the parameter is moved according to a percentile range obtained out of a historical distribution or a conservative set of parameters or simply by taking a multiple of the portfolio sensitivity to that parameter. Calibration uncertainty can be estimated varying calibration sets and model risk by comparing valuation of the same deal with different models. Finally, the simulation of hedging strategies can also provide a good estimation of model risk as seen for instance in \cite{Elices2011}. The main disadvantage of this approach is the big computing capability needed which may be impractical at portfolio level.

\begin{figure}[htbp]
	\centering
	\includegraphics[width=0.99\textwidth]{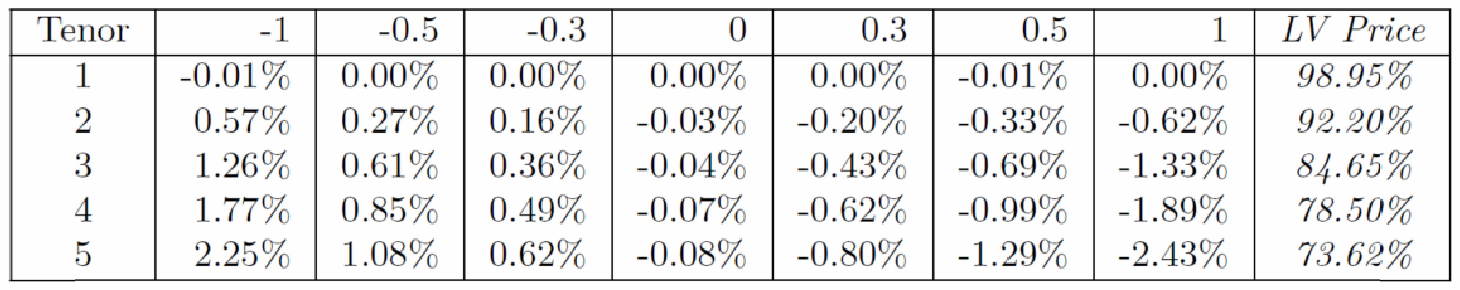}
	\caption{LV - HWLV prices, varying maturity (tenor) and equity-rates correlation.}
	\label{tab:SampleFigure}
\end{figure}

Table \ref{tab:SampleFigure} shows an example of an FVA accounting for stochastic rates on an autocallable product on STOXX50E with market data taken from June 10th, 2012. It shows the premium difference of Hull-White model with local volatility (HWLV) minus local volatility model (LV), varying maturity (tenor rows) and correlation between equity and rates (each column). The product cancels yearly when the underlying return is above 100\%, paying the notional back plus a yearly increasing coupon of 5\% the first year, 10\% the second and so on. While alive, it pays floating coupons and at maturity the client receives a 100\% redemption and pays a put striked at 50\% and a digital put with leverage 50\% and strike 50\%. With current levels of historical correlation around 0.3, the HWLV model can be 80 basis points more expensive than LV for a 5 year maturity deal.

The FVA can be calculated externally or sometimes it may be incorporated inside the premium of each deal. For instance, a conservative correlation instead of the market one, embeds the FVA inside the deal and it does not need external calculation (taking the difference between conservative and market correlations). Another example arises with maximum delta or gamma softening. A conservative softening embeds an FVA by reducing delta and gamma. This increases the deal price to account for liquidity premium and helps the management of the position. External FVA helps avoiding collateral disputes and allows better control tracking, whereas an embedded FVA usually helps position hedging and management.

It is not a good practice to embed a higher FVA for wrong concepts. This happens for instance when an excess of softening is applied to account for forward skew instead of liquidity. The advantage of this policy is that it helps the management of the position and hedges a more expensive payoff which gets recovered on average. However, the uncertainty of the hedging error will be higher, because the increase of premium is not a consequence of an actual modelled source of risk.

\section{Conclusions}
\label{sec:Conclusions}

The current trend of pricing models evolves towards more complexity and simpler products. Modelling improvements are less about developing new models and more about improving existing ones, their inputs and outputs and their integration into a well coordinated pricing and risk management system. The model validation function comes into play to guarantee the permanent improvement of pricing models, the correct management and mitigation of model risk and making senior management and model users aware of model risk.


The paper addresses good practices for management and mitigation of model risk and ways to calculate fair value adjustments. A worked example on the impact of stochastic rates on autocallable products is provided for illustration purposes.

\paragraph{Acknowledgements:} The author wants to thank Pablo Blanco, Eulogio Cuesta and Peter Walsh for their contributing ideas and discussions and Diana Gon\c{c}alves and Santander Front Office Development Group for the worked example.

%


\begin{thebibliography}{30}

\bibitem{Cont2006}
Cont R., ``Model uncertainty and its impact on the pricing of derivative instruments.'', \emph{Mathematical Finance}, Vol. 16, No. 3, pp. 519-547, 2006.


\bibitem{Elices2011}
Elices A., Gimenez E., ``Applying hedging strategies to estimate model risk and provision calculation'', to appear in \emph{Quantitative Finance}, 2012 available at ``http://arxiv.org/abs/1102.3534''.

\bibitem{Morini2011}
Morini M., ``Understanding and Managing Model Risk: A practical guide for quants, traders and validators'', \emph{Finance series, John Wiley and Sons}, 2011.


\end{thebibliography}
\end{document}